*This is the authors' version of the work.*
*This is a preprint of an article published in PACIS 2021 Proceedings*

# Preprint Notes

**Title:** Towards a Trustful Digital World: Exploring Self-Sovereign Identity Ecosystems

**Authors:** Gabriella Laatikainen, Taija Kolehmainen, Mengcheng Li, Markus Hautala, Antti Kettunen and Pekka Abrahamsson

# TOWARDS A TRUSTFUL DIGITAL WORLD: EXPLORING SELF-SOVEREIGN IDENTITY ECOSYSTEMS

*Completed Research Paper*

**Gabriella Laatikainen**
Faculty of Information Technology
University of Jyväskylä, Finland
gabriella.laatikainen@jyu.fi

**Taija Kolehmainen**
Faculty of Information Technology
University of Jyväskylä, Finland
taija.s.kolehmainen@jyu.fi

**Mengcheng Li**
Faculty of Information Technology
University of Jyväskylä, Finland
mengcheng.m.li@jyu.fi

**Markus Hautala**
TietoEVRY
Espoo, Finland
markus.hautala@tietoevry.com

**Antti Kettunen**
TietoEVRY
Espoo, Finland
antti.j.kettunen @tietoevry.com

**Pekka Abrahamsson**
Faculty of Information Technology
University of Jyväskylä, Finland
pekka.abrahamsson@jyu.fi

## Abstract

*In the current global situation—burdened by, among others, a vast number of people without formal identification, digital leap, the need for health passports and contact tracking applications—providing private and secure digital identity for individuals, organizations and other entities is crucial. The emerging self-sovereign identity (SSI) solutions rely on distributed ledger technologies and verifiable credentials and have the potential to enable trustful digital interactions. In this human-centric paradigm, trust among actors can be established in a decentralized manner while the identity holders are able to own and control their confidential data. In this paper, we build on observations gathered in a field study to identify the building blocks, antecedents and possible outcomes of SSI ecosystems. We also showcase opportunities for researchers and practitioners to investigate this phenomenon from a wide range of domains and theories, such as the digital innovation ecosystems, value co-creation, surveillance theory, or entrepreneurship theories.*

**Keywords:** Decentralized identity, Self-sovereign identity, Innovation ecosystems, Digital trust

## Introduction

In the current global situation, the need to provide individuals and other entities a secure and private digital identity has been accentuated. More than a billion people lack a formal identity that would give them access to a variety of services, such as financial services, education, healthcare and several aid programs (Ladner et al., 2016). The refugee crisis in Europe left several people without formal documents, causing data management challenges in recipient countries, such as privacy and security issues of centralized databases that store sensitive biometrics and biographical data (Lemieux, 2017).

The pandemic crisis has accelerated the digitalization and forced individuals, organizations and governments to use digital systems without careful considerations of the users' data security and privacy (Morley et al., 2020). We live our lives by leaving our rapidly growing digital footprints without giving informed consent for the (re-)use of our data in all cases (Cinnamon, 2017). In fighting against infectious disease outbreaks, governments and other authorities face challenges in assessing the ethical justifiability of digital applications that use sensitive users' data to stop the spread of the virus and ensure privacy, fair data sharing, responsible data use, discrimination, freedom of movement and voluntariness (Morley et al., 2020; Sharon, 2020). Indeed, the question of how to identify individuals, what data to collect about them, where these data should be stored, who owns and controls the data, is a complex, timely, and important matter.

Self-sovereign identity (SSI) solutions promise a trustful, private and secure way of digital identity management, offering new insights into the above mentioned problems (Mühle et al., 2018). In currently used centralized and federated identity management systems services, organizations and platform providers provide the users various digital identifiers and identification means and require the users to trust their systems (Mühle et al., 2018; Grüner et al., 2019). The concept of SSI relies on the concepts of decentralized identity that take advantage of distributed ledger technologies (a broader class of "blockchain-inspired" technology; Zachariadis et al., 2019) and verifiable credentials (i.e., a tamper-evident credential that has authorship and can be cryptographically verified; W3C, 2020) to provide the users digital identity in a decentralized manner without relying on intermediaries (i.e., trusted third parties; Naik and Jenkins, 2020). As a distinction to decentralized identity systems, the SSI paradigm has additional requirements that ensure the users' sovereignty of their identity and the storage-control of the associated confidential data linked to their identity (Naik and Jenkins, 2020) and thus, might offer a new digital identity infrastructure for the post-industrial age.

Besides providing digital identity for individuals, SSI systems have the potential to create new and transform existing organizational processes in a variety of industries by providing digital identity to other entities, such as organizations and things (e.g., IoT devices or cars; Lemineux, 2017). For example, apart from reforming the authentication and authorization processes, SSI systems enable proving the steps of a journey of physical and digital things in a supply chain; they might eliminate the privacy and identification concerns related to patient data in digital healthcare systems, and might establish trustful value transfer and communication with partners, customers and regulators in business ecosystems (Zwitter et al., 2020). SSI systems can also enable the emergence of new business models, such as the identity insurance schemes, and the emergence of value-stable cryptocurrencies ("stablecoins") that function as local currencies (Wang and De Filippi, 2020).

Despite their potential benefits, transforming existing processes and orchestrating new digital identity ecosystems carry several challenges, such as the fragmentation of the SSI market, lack of standards and regulations, the immaturity of the technology, legal uncertainty, and challenges related to decentralized governance (Lemieux 2017; Wang and De Filippi, 2020; Prewett et al., 2020). These challenges question the feasibility of the SSI ecosystems and as such, require further investigations.

Due to the nascent nature of SSI technology and their ecosystems, our current understanding of the phenomenon is limited. There are a few studies investigating the technology aspects of SSI (e.g., Mühle et al., 2018; Naik and Jenkins, 2020; Dunphy and Petitcolas, 2018), user requirements (e.g., Ostern and Cabinakova, 2019), trust requirements (e.g., Grüner et al., 2020), philosophical and legal perspectives of identity (Zwitter et al., 2020) and cases of the real-world adoption of SSI (e.g., Wang and De Filippi, 2020). However, there is a need for studies that would lay the foundations for theory development and advance our understanding on blockchain-related technologies and related concepts (Rossi et al., 2019). In particular, SSI requires investigation from a holistic view (Gstrein and Kochenov, 2020) and we lack studies that take a more comprehensive view and study SSI from an ecosystem perspective.

To address this gap, in this study, we describe the results of a field study carried out in close collaboration with (1) a partner company aiming to orchestrate a digital identity infrastructure, as well as (2) several domain experts who are contributing members in a standard setting organization (i.e., the TrustOverIP Foundation; ToIP, 2021). We aim to answer the following research question: *"How can SSI ecosystems be defined, and what are the antecedents and possible outcomes of their orchestration and adoption?"*. We base our results on our observations as well as interviews with key domain experts.

The contribution of this study is two-fold. First, we aim to narrow the gap between researchers and practitioners and propose a model that describes the building blocks of SSI ecosystems, their potential outcomes, as well as the factors that have an impact on their characteristics, orchestration and possible adoption (i.e., antecedents). Second, we propose further research avenues from the viewpoint of several research domains and theories that might facilitate the adoption and orchestration of SSI ecosystems.

## Self-Sovereign Identity Ecosystems

SSI is an emerging concept that can be viewed as (1) an identity management system, (2) a human-centric data management paradigm or (3) an identity protocol. First, as opposed to the central identity management system, SSI is a decentralized identity management system that enables individuals and other entities to manage their identity and personal data associated with their identities by storing them locally on their own devices (or remotely on a distributed network) and selectively give access to authorized third parties, without the need to refer to any trusted intermediary to provide or validate these claims (Mühle et al., 2018). Second, SSI can be viewed as a human-centric data management paradigm where the users own and control their identity and the personal data associated to their identity (Naik and Jenkins, 2020). Third, SSI can be considered as an identity protocol, a commodity providing private, secure and trustworthy data storage and communication (Zwitter et al., 2020).

Although the concepts of SSI and decentralized identity are typically used interchangeably, there is no consensus in the literature and among experts on the definition of SSI. However, there is a general understanding that SSI "is intended to preserve the right for the selective disclosure of different aspects of one's identity and the various components thereof, in different domains and contextual settings." (Wang and De Filippi, 2020, page 9). In other words, SSI refers to decentralized identity with additional requirements that can be used to assess SSI solutions. Self-sovereign identities should be secure (i.e., the identity must be kept secure), controllable (i.e., the user must be in control of who can view and access their data), and portable (i.e., the user must be able to use their identity data wherever they want and the data should not be tied to a single provider) (Allen, 2016). It should be noted that the assumption that an SSI solution fulfils all requirements, is not fully plausible (Wang and De Filippi, 2020).

There are several non-profit organizations, open source communities and standard setting organizations [1] aiming to define, standardise and provide tools for the decentralized identity architectures and the digital interactions that these enable. Despite the existence of laws and regulations related to digital identification, data exchange and protection (e.g., eIDAS, Pan-Canadian Trust Framework, GDPR) there is still legal and regulatory uncertainty in the global market. While there is currently no global standard for SSI implementations, the most prevalent elements of SSI technology is proposed by the World Wide Web Consortium [2] (W3C, 2020) and the Decentralized Identity Foundation [3](DIF, 2021). The concept of SSI Ecosystem has been promoted by a standard setting organization, the TrustOverIP Foundation [4] (ToIP, 2021). While other decentralized identity communities concentrate on some of the aspects of SSI systems (e.g., technological) or some of the industries (e.g., healthcare), this standard setting organization focuses on "internet-scale" (i.e., cross-country and pan-industry) SSI solutions and the ecosystems built around the technology.

According to W3C and DIF standards, the key components of currently developed decentralized identity management systems are the decentralized identities (DIDs; DID, 2021) and a secure, private and encryption-based communication protocol called DIDComm (DIDComm, 2021). DIDs are URL-based identifiers ensuring the portability of the credentials without the need to reissue the credential (DID, 2021). They are used in combination with verifiable credentials and verifiable claims that provide the possibility to make any number of attestations about a DID subject that grant access to rights and privileges to trusted authorities (Dunphy and Petitcolas, 2018). The DIDComm specifies the communication between DIDs such as connecting and maintaining relationships and issuing credentials, providing proof, and its design is aimed to be secure, private, interoperable, transport-agnostic and extensible (DIDComm, 2021; Windley, 2020). DIDComm has been supported by libraries, agents, applications implemented in different languages, such as Python, .NET, GoLang (DIDComm, 2021).

[1] These include, for example, the World Wide Web Consortium, Decentralized Identity Foundation, Internet Engineering Task Force (IETF), TrustOverIP Foundation, MyData, Sovrin, OASIS, OpenID, FIDO Alliance, Hyperledger, Open Identity Exchange, CULedger, Alastria, uPort, Civic, SelfKey, Alastria, and Global ID.
[2] The W3C is supported by major internet and technology companies, universities and governments, such as Amazon, Apple, Boeing, Cisco, Microsoft, Google, Facebook, Alibaba, Tencent, and Baidu.
[3] The DIF is supported by most of major blockchain identity and data companies, such as Microsoft, IBM, Hyperledger, Accenture, Mastercard, RSA, Civic, uPort, BigChainDB and Sovrin.
[4] The TrustOverIP Foundation was founded in May 2020 and aims to combine the open standards, architectures, and protocols developed in other standard setting organizations and technical development partners. Currently, it has more than 150 members, with steering members such as IBM, British Columbia, Accenture, Evernym, Finicity, CULedger, Futurewei, LG CNS and IdRamp.

SSI solutions are typically built and managed by a collaborative effort among different actors forming an ecosystem (Wang and De Filippi, 2020). Technically, digital identification can be established through the interaction of three roles that together form "the digital trust triangle" (ToIP, 2021; Davie et al., 2019): issuers, holders and verifiers. *Issuers* are the source of credentials; they determine what credentials to issue, what the credential means, and how they validate the information assigned to the credential. *Holders* request credentials from issuers; they hold and present them when requested by verifiers and approved by the holder. Holders can be individuals, organizations, or other entities. *Verifiers* request the credentials they need and then follow their own policy to verify their authenticity and validity. This "triangle" is managed by a governance authority that may represent any set of issuers organized in different forms (for example, government, consortia, cooperative). The role of governance authorities is to publish a governance framework that consists of rules for managing the ecosystem, in particular the business, legal, and technical policies for issuing, holding, and verifying the credentials. As an example, in a payment card ecosystem (e.g., Mastercard), the governance authority is Mastercard, the issuers are the banks and credit unions, the holders are the individuals or businesses applying for Mastercard, and the verifiers are merchants enrolled in the Mastercard ecosystem to accept payment cards. (ToIP, 2021)

In SSI ecosystems, the trust among actors is established through peer-to-peer interactions assured by (1) the technology that provides immutable and secure data storage, data exchange and communication and by (2) the governance frameworks consisting of business, legal, and technical rules and policies (Davie et al., 2019; ToIP, 2021). Thus, SSI solutions refer not only to a technological solution, but also require a governance framework, and these two components are interrelated (see, e.g. Zwitter et al., 2020). Thus, in this study, we chose SSI ecosystems as the unit of analysis, which allowed us to study SSI from a holistic perspective.

## Research Method

The emergence of SSI ecosystems is a rather new phenomenon where changes in technology and in the market happen rapidly. To bridge the gap between practitioners and researchers, we chose an exploratory qualitative approach that is capable to encompass empirically rich and detailed data related to a nascent complex phenomenon based on human actions (Edmondson and McManus, 2007; Myers, 2009). We gathered the data by conducting field research. Field research involves collecting data in form of variable methods, such as direct observation, participation in the life of the community, interaction with community members, collective discussions and analyses of documents produced within the community. (Myers, 2009)

The study was carried out between April 2020 and March 2021. In this study, we collaborated closely with an IT service provider company in form of a research project. This company has been developing a decentralized identity infrastructure in collaboration with other private and public organizations. As a part of this work, we joined a standard setting organization, the TrustOverIP Foundation (for a short description, please refer to section Self-Sovereign Identity Ecosystems, footnote 4; ToIP, 2021) as contributing members. Being an active member of the community made it possible to connect with several key domain experts and observe the SSI ecosystems in real context. During the field work, we focused on investigating the main characteristics of these ecosystems, the drivers and challenges of their development and adoption, and the behavioural foundations of the contributing experts.

### *Data collection*

The data for this research has been collected from several sources. First, we actively collaborated with our partner company in form of project meetings, emails and informal discussions related to orchestrating a particular decentralized identity ecosystem. Second, one of the authors actively participated in TrustOverIP meetings, email conversations, and informal discussions related to the following tasks: (1) contributing to the human experience task force, and (2) creating a survey for TrustOverIP community to understand the current state-of-practice of SSI adoption, benefits, and barriers. Third, to advance our understanding of the phenomenon, we participated in webinars and university lectures given by domain experts. Fourth, we actively followed news, social media postings, and the conversations of the TrustOverIP community via Slack. We collected insights from the websites of several companies, non-profit organization, standard setting organizations that are relevant in this domain. The most important events are listed in Table 1; however, the authors have had earlier experience with other blockchain-related research and empirical work that are not included as a form of data collection for this particular study.

**Table 1. List of events.**

| Events | Duration |
|---|---|
| Project meetings with our collaborative partner and research group meetings | 36 meetings with an average of 60 minutes |
| Active participation in TrustOverIP meetings | 5 meetings of cca. 60 minutes |
| Active participation on webinars and university lecture given by key domain experts | 6 webinars/lectures sum. cca. 440 minutes |
| Summary | 50 hours 20 minutes |

We systematically gathered our observations for further analysis in the form of field notes in several ways. First, we listed our observations in the format of date, source, person, and observation/key insight. Second, we took screen captures about important presentations and meetings. Third, we recorded some relevant parts of the meetings with the consent of the participants. Fourth, we extracted relevant information from Slack chat records, the chat history of the webinars, lectures, presentation slides and websites.

Working together with field experts and observing their work closely allowed us to understand the phenomenon in general. However, we wanted to get insights into some more specific questions in depth, and thus, as an additional data source, we carried out ten in-depth interviews with experts from The United States of America, Canada, and Finland. These interviews followed an open-ended interview structure (Darke et al., 1998) and their length varied between 36–68 minutes. These were two group interviews and eight individual interviews. All interviews were recorded and transcribed. Table 2 provides an overview of the informants.

**Table 2. List of informants.**

| Interviewee | Organization/Institution Type | Position at the Organization/ Institution | Other affiliations | Number of interviews |
|---|---|---|---|---|
| Interviewee 1 | IT Service Provider, Organization A | Senior Blockchain Consultant | MyData[5] member | 2 group interviews and 1 individual interview |
| Interviewee 2 | IT Service Provider, Organization A | Head of Innovation Center | MyData member | 2 group interviews and 1 individual interview |
| Interviewee 3 | Public Digital Trust Service Provider (government), Institution B | Executive Director | Executive Director of the TrustOverIP Foundation | 1 group interview |
| Interviewee 4 | Provider of Platform for Verifiable Credentials, Organization C | Chief Trust Officer | TrustOverIP Steering Committee Member, W3C DID Co-editor, Sovrin[6] Co-Chair, OpenID Foundation[7] Founding Board member, Identity Commons[8] Steward | 1 group interview |

[5] https://mydata.org/
[6] https://sovrin.org/
[7] https://openid.net/
[8] https://www.idcommons.org/

| | | | | |
|---|---|---|---|---|
| Interviewee 5 | Innovative Consultant Services Provider, Organization D | Founder and CEO | TrustOverIP member | 2 individual interviews |
| Interviewee 6 | Innovative Consultant Services Provider, Organization E | Consultant and Advisory | TrustOverIP member, DIF contributor, Identity Defined Security Alliance[9] Member, Sovrin Board Member | 1 individual interview |
| Interviewee 7 | - | - | TrustOverIP member, Sovrin member | 1 individual interview |
| Interviewee 8 | Certified Public Accountant, Cybersecurity Audits Organization F | Cybersecurity Consulting Audit and Governance Executive | TrustOverIP Working Group Co-Chair, Sovrin member, CULedger Board of Advisors member, Credential Master, Board of Advisors member | 1 individual interview |
| Interviewee 9 | Service and Infrastructure Provider, Bank | Principle Technology Strategist | MyData member | 1 individual interview |

Our interviewees were very active, contributing members in one or more standard setting organizations or other non-profit communities working in digital identity domain. Further, they have had work experience in digital identity domain for several years in distinct roles. Moreover, two of the interviewees were founders of the TrustOverIP Foundation. The diversity of the background of our interviewees made it possible to gather insights from different viewpoints. The interview questions were personalized around the viewpoints and key themes enlisted in Table 3.

**Table 3. Themes for the interviews.**

| Viewpoint | Key themes |
|---|---|
| Open source communities and standard setting organizations | – SSI ecosystems in general (e.g., definition, key terms)<br>– The roadmap towards mass adoption<br>– Challenges in the journey towards mass adoption<br>– Possible indicators of success of the TrustOverIP Foundation<br>– Reasons for founding the TrustOverIP Foundation |
| Organizational perspective | – Organizational factors impacting the SSI ecosystems<br>– The perceived risks and benefits of SSI ecosystems<br>– The key barriers in orchestrating and adopting SSI ecosystems<br>– How customers and other partner companies perceive the value |
| Individual perspective | – Personal incentives to work in SSI domain and contribute<br>– Personal objectives<br>– The success indicators of the work related to SSI ecosystems |

Before the data analysis, we combined the data coming from different sources (field notes, interview transcriptions, etc.). We extracted the relevant information collected in video and picture formats into concise text. As a result, we had cca. 110 pages of relevant data.

## *Data analysis*

In the data analysis, we used open coding, axial coding and theoretical coding to analyze the data systematically in an iterative manner. In the *open coding* phase, we summarized different parts of the text using succinct codes by applying the constant comparative method (Locke, 2002; Myers, 2009). We concentrated on identifying the aspects that appeared to have relevance to the phenomenon. In this phase, the coding was concrete, detailed and in most cases, followed the wording of the original text (Strauss and Corbin, 1994). This phase resulted in several codes, such as "trust in government", "feasibility", "interoperability", "data policies", "digital inclusion" and "self-esteem". In the *axial coding* phase, we focused on finding regularities, patterns, explanations, and causalities between the codes. Based on our observations, we assigned the codes into broader categories using tables that represented different aspects of the phenomenon, such as "Organizational and institutional outcomes", "Collaborative ecosystems" and "Industry characteristics". In the *theoretical coding* phase, we focused on formulating a model that provides researchers and practitioners a holistic view of the phenomenon and helps to identify further research gaps (e.g., by focusing on some of the items, or, combining specific items and their interrelations). After some iterations, we found that the Antecedents-Concept-

[9] https://www.idsalliance.org/

Outcome model fits the data and the purpose of the analysis very well. Thus, we assigned the categories into one of the following groups: (1) Building Blocks, (2) Antecedents and (3) Possible outcomes. Lastly, we carried out several discussions related to this model and its items both in our research group as well as with experts from our collaborative firm. In all phases of data analysis and validating discussions, we renamed, relocated, merged, deleted the codes several times to eliminate the illogicalities and repetitions. In the end, all authors agreed on the proposed model.

## SSI Ecosystems: Building Blocks, Antecedents and Potential Outcomes

As a result of our study, we propose a model that describes the building blocks, antecedents, and possible outcomes of SSI ecosystems. In Figure 1 this model is shown with the categories and underlying codes that resulted from our data analysis. In the next three subsections, we describe these items in more detail; however, due to page limitations, we also put the results into context in the same subsection by referring to recent literature.

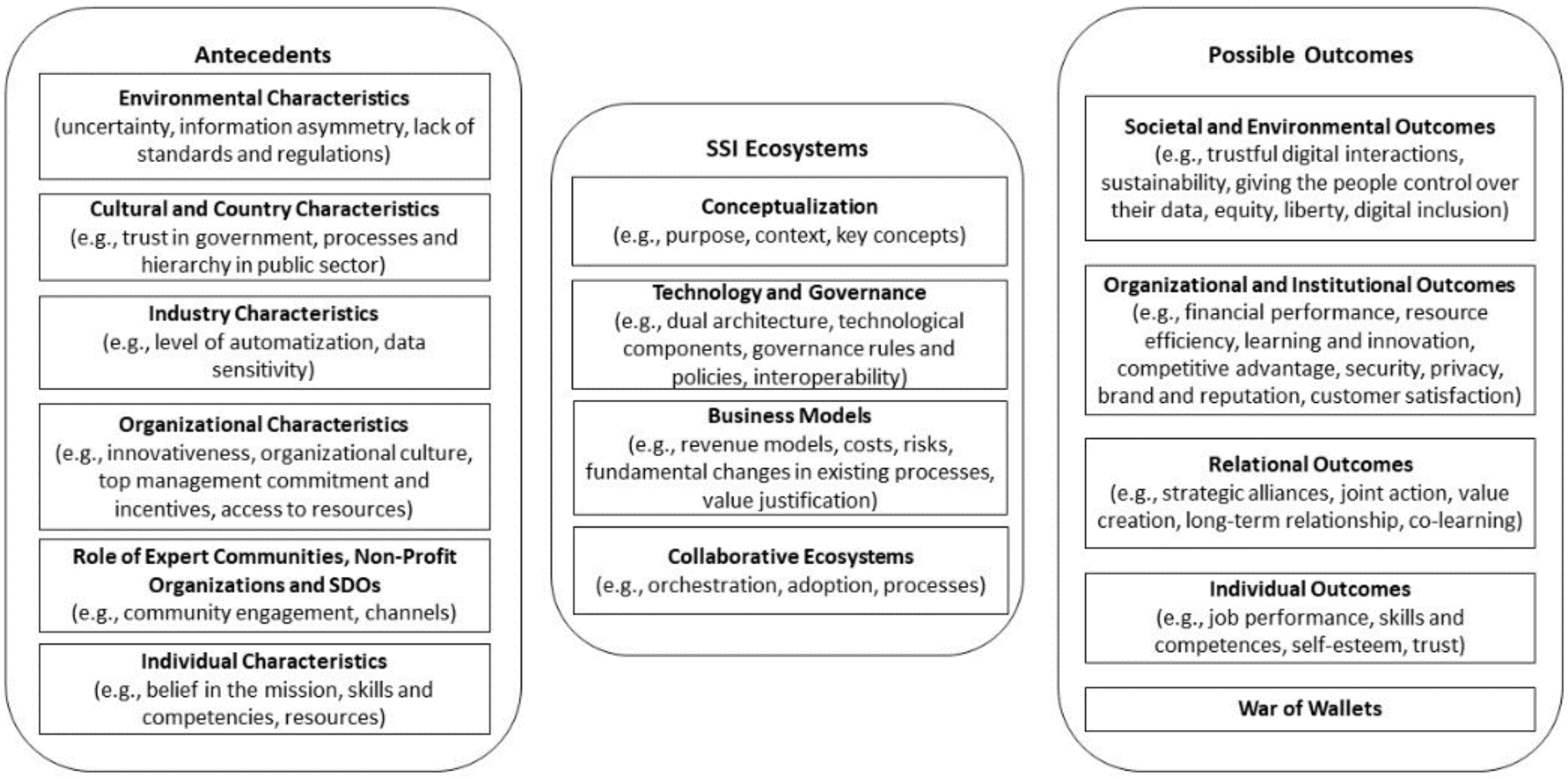

**Figure 1. SSI Ecosystems: Building Blocks, Antecedents, and Possible Outcomes.**

### *Building blocks of self-sovereign identity ecosystems*

We build on the definition of innovation ecosystems by Moore (1996) and define SSI Ecosystems as innovation ecosystems describing the collaborative effort of a diverse set of actors towards innovation. In these ecosystems, the key actors are the infrastructure providers delivering a technology and governance dual framework and various partner companies and end customers that provide complementary products and services as well as demands and capabilities. Based on our investigation, we argue that SSI Ecosystems consist of four components that need further elaboration when orchestrating or adopting a particular SSI ecosystem in practice. First, there is a need for proper *conceptualization*, such as defining the purpose, mission, context and boundary conditions. Second, there is a need for SSI infrastructure that consists of *a technology and governance architecture*. Third, SSI ecosystems need carefully crafted *business models* that provide fair value for each actor. Fourth, the orchestration and adoption of *collaborative ecosystems* requires designing and managing the roles, processes and activities and shaping an innovative culture and environment. In what follows, we describe these building blocks in more details.

**Conceptualization:** SSI ecosystems need proper conceptualization, which can be achieved by defining their purpose and context, launch style and formative ideology (van Pelt et al., 2019; Hsieh et al., 2017). Digital identity in general, might refer to "all attributes of a person that uniquely defines the person", whereas personas are specific facets of an identity that is expressed in a particular context (Wang et al., 2020, pg. 2). The same persons might hold multiple personas in different contexts. Thus, orchestrating an SSI ecosystem requires not only defining the digital identity and the personas in this particular context, but also the required levels of digital trust and trust assurance (ToIP, 2021).

**Technology and Governance Dual - Framework:** SSI solutions cannot be created without proper considerations of governance decisions (Zwitter et al., 2020). Thus, SSI solutions require both technology and governance frameworks that are interrelated and have to be defined side by side. When building digital identity ecosystems, the first step is to collect the business and policy requirements, and only then it is possible to choose the technologies and the components to implement these (ToIP, 2021). There are various proposed architectures under development that need further assessment (e.g. Civic, uPort). One possible SSI architecture has been proposed by Davie et al., (2019) that is developed and promoted by the TrustOverIP Foundation as a future standard (ToIP, 2021). This architecture consists of both technology and governance frameworks divided into four layers: the first two layers are responsible for technical trust (machine-to-machine) and the second bottom layers represent human trust (ToIP, 2021).

**Business Models:** The business model of SSI ecosystems needs careful design and communication. The adoption of SSI solutions offers several benefits; however, it requires fundamental changes in existing legacy systems and business processes (Wang and De Filippi, 2020). Thus, the currently used business models might need to be fully redesigned, for example, by requiring transaction fees for credential use. Crafting a value proposition at the ecosystem level is very challenging and needs to consider different, mostly contradictory objectives of the actors (De Filippi and Loveluck, 2016). Another challenge is related to value justification, i.e., communicating the value to different actors in the ecosystem (see e.g., Töytäri et al., 2011). The potential benefits of SSI are typically visible only after a long time period, whereas adopting SSI requires significant start-up investments and taking high risks due to, for example, regulatory and legal uncertainty, lack of standards, lack of knowledge and skills and immature technology (Zwitter et al., 2020).

**Collaborative Ecosystems:** SSI ecosystems consists of a set of different type of actors, such as private and public organizations, non-profit organizations, service providers, business customers and individual customers. SSI ecosystems are typically collaborative ecosystems where achieving good governance is especially challenging (Zachariadis et al., 2019; Lacity, 2018). The roles, rules and rights of the actors should be defined carefully in order to provide incentives and advance the actors' engagement (van Pelt et al., 2019). Good governance can be defined and achieved in multiple ways (e.g., effective in decision making, conflict management, adaptation and change, effective in providing incentives; protecting the founders' investment; Hsieh et al., 2017). In the forming/genesis phase, the role of a "hidden leader" has been accentuated (Lacity et al., 2019). In the operating phase, another challenge is related to the intention of collaborative actors that have impact in the ecosystem without having formal authority (Storbacka et al., 2016).

## *Antecedents*

The characteristics, adoption and orchestration of SSI ecosystems are affected by several factors. In what follows, we describe these antecedents in more details.

**Environmental characteristics:** SSI ecosystems reside in a highly uncertain environment characterized by high volatility, fast changes, and information asymmetry. The fragmented nature of SSI market, the immaturity and incompatibility of standards, legal and regulatory uncertainty have an impact on SSI ecosystems and their possible adoption.

**Cultural and country characteristics:** Cultural and country characteristics are key antecedents of SSI ecosystems. For example, the potential benefits of an SSI solution aiming to achieve digital and financial inclusion are higher in developing countries; however, this implies additional challenges (e.g., citizens do not have smartphones to store their private keys securely; Wang and De Filippi, 2020). Furthermore, the diversity among countries related to, e.g., the citizens' trust in governments, the processes and hierarchy in public sector, affects the individuals' adoption behaviour and the requirements of the SSI solution.

**Industry characteristics:** Industry characteristics, such as the level of automatization and required data sensitivity, influence the potential benefits and the associated risks of SSI ecosystem adoption. For example, the potential benefits in healthcare, insurance, or education might outweigh the required sacrifices, whereas in other industries, the perceived value of SSI ecosystem adoption remains too low.

**Organizational characteristics:** Organizational characteristics, such as innovativeness, the organizational culture, the commitment and incentives of top management and the access to resources have a high impact on the decisions made related to SSI ecosystems.

**Role of expert communities, non-profit organizations and standard setting organizations:** Open source communities, non-profit organizations and standard setting organizations play an important role in orchestrating SSI ecosystems. These communities offer a possibility to learn, get to know new people, obtain reputation and a brandname, and they can also increase the members' self-esteem and feeling of usefulness. The digital identity community collaborates actively via different channels, such as meetings, Slack discussions, emails, social media posts, webinars and conferences.

**Individual characteristics:** Individual characteristics, such as the skills, beliefs and resources of citizens or other individuals play an important role when studying the adoption of SSI ecosystems. Similarly, experts' skills, networks, and commitment in organizational or community settings strongly influence the orchestration of these ecosystems.

### *Possible outcomes*

SSI ecosystems offer a wide variety of outcomes that we describe in more details next.

**Societal and symbolic outcomes:** SSI relies on a human-centric data management philosophy where the user is at the centre and there is no need for third parties to issue and administer an identity (Dunphy and Petitcolas, 2018). Thus, the perceived value of these ecosystems has a societal and symbolic dimension. SSI promises to provide a fundamental change in the digital world by making digital interactions trustful (ToIP, 2021). This paradigm presents sustainability, equity, freedom, liberty and digital inclusion when these ecosystems are adopted for different use cases, such as healthcare, digital voting and digital education (e.g., Zwitter et al., 2020; Houtan et al., 2020).

**Organizational and institutional outcomes:** Adopting SSI is expected to simplify and automatize organizational processes, achieve cost savings, increase customer satisfaction and the reputation of the company. Furthermore, SSI can open up new business opportunities by developing innovations in form of new services, processes or business models. Thus, making digital trust a priority by adopting SSI is perceived as a new way to achieve competitive advantage in the market and stimulate customer retention, among others.

**Relational outcomes:** SSI ecosystems provide an opportunity to form new strategic alliances and facilitate new ways of value creation and joint actions. Orchestrating and adopting SSI ecosystems enable networking, co-learning and building of long-term relationships among the actors.

**Individual outcomes:** SSI ecosystems have the possibility to impact all our lives by providing a private and secure way of digital interactions. Their adoption might lead to higher job performance and trust in digital systems. Furthermore, orchestrating SSI ecosystems might result in passion at work, new knowledge and skills, better self-esteem and a sense of usefulness, among others.

**War of Wallets:** The great technology companies (e.g., Apple, Microsoft) are also interested in developing SSI solutions (Forbes, 2020) that might further consolidate their power and influence. SSIs reportedly have the power to determine the future of the Internet by providing an opportunity to control what you can do, where you can go, and who you can be in your digital life (Reed, 2020). The War of Wallets–similarly to the War of Browsers (see, e.g., Wikipedia, 2020)— refers to a possibility of an impending war among different players on who will control our data (Reed, 2020; Evernym, 2020).

## Self-Sovereign Identity Ecosystems: Further Research Avenues

Being a nascent concept, to achieve some of the potential benefits, SSI ecosystems require further investigation of a wide range of complex research issues from different viewpoints and theories. The elements of the model in Figure 1 as well as their interrelations suggest several intriguing research opportunities that we will describe next. First, the emergence of SSI widens our understanding on several *concepts* and creates a need to reinvestigate these. For example, IS researchers could focus on studying the concept of digital identity in different contexts (e.g., self-identity, personal identity, organizational identity, social identity; Whitley et al., 2014; Zwitter et al., 2020). Furthermore, researchers might focus on (re)defining SSI, digital identity, digital trust and related concepts (such as trusted interaction, trusted requirement, implicit trust, trusted processes, trust enablement, transitive trust and chain of trust), by studying, for example, the interplay of digital identifiers and personas (Wang and De Filippi, 2020) or the role of SSI in embedding trust into digital interactions (Grüner et al., 2020).

There is also a need for research related to *technical aspects* of SSI ecosystems. The technical requirements, the architecture, design principles, testing suites, technical interoperability and abstracting interfaces that allows comparison, assessment and management of different solution architectures need to be elucidate (e.g., ToIP, 2021; Naik and Jenkins, 2020; Coelho et al., 2018). Researchers interested in data semantics could focus on "decentralized semantics", i.e., how a data object within distributed ledger technology solutions represents a concept or object in the real world (ToIP, 2021). Related to the *governance framework* of SSI ecosystems, research is needed to investigate the governance architecture requirements, design principles, roles, responsibilities of actors, governing processes, the levels of trust, levels of assurance and the risks related to digital identity and verifiable credentials (ToIP, 2021). The horizontal interoperability of different ecosystems—specifying the architectural dependencies and requirements as well as the specific inner and joint governmental processes needed for interaction—is yet to be developed (ToIP, 2021). SSI creates an opportunity for innovative ideas and new markets around digital identity (e.g., identity insurance; Wang and De Filippi, 2020). It also requires creating new *business models* and redefining the existing ones. Thus, there is a need for further research on the business aspects of SSI ecosystems, such as the revenue models and cost structures.

More research is needed on the *adoption* of SSI ecosystems from different viewpoints. Researchers might focus on the feasibility, facilitating conditions, drivers, barriers, success metrics, and the impact of these ecosystem at different levels (individuals, organizations, industries, countries, etc.). Their market potential and the factors affecting mass adoption should be studied as well (Ondrus et al., 2015). Qualitative studies on SSI solutions from *different domains* (e.g., healthcare, education, IoT, online voting, supply chain, etc.) are also needed to increase our understanding on the diversity of aspects in SSI research area (ToIP, 2021; Zwitter et al., 2020). The emergence of SSI provides new aspects to study the *digital and financial inclusion* (Wang and De Filippi, 2020). Researchers could address the cultural aspects and the difference in the mental models, for example, from the decolonizing design theory perspective (e.g., Tlostanova 2017).

One of the challenges that needs further research is related to *ecosystem orchestration and operation*. IS researchers focusing on ecosystems and collaborative economies could advance our understanding by investigating several issues, such as (1) factors impacting operational decision making (Muñoz and Cohen 2018), (2) roles in the ecosystem and the multilateral nature of relationships between them (Dedehayir et al., 2018; Kapoor, 2018), (3) processes for ecosystem orchestration and operation (e.g., Jacobides et al., 2018) and (4) the sustainability of these ecosystems (e.g., Sussan and Acs, 2017). Further research is needed also related to define the role of "de facto leader" of these ecosystems, for example, from microfoundational perspectives (Storbacka et al., 2016). Governance issues, such as the boundary problem of defining the actors (Allen and Berg, 2020), the accommodation of incompatible and often irreconcilable interests and values (De Filippi and Loveluck, 2016), the co-existence of informal and formal contracts (Allen and Berg, 2020), the interplay between machine readable and traditional governance mechanisms (Lumineau et al., 2020) and the dynamics of governance mechanisms and their mutual influence over time (Lumineau et al., 2020) should also be studied.

There is a need to address SSI also from the *value co-creation* perspective. For example, researchers could study the different facets of value that these ecosystems enable, how different actors perceive it, and the co-creationary processes, such as resource integration and social interactions (Li and Tuunanen, 2020; Rajala et al., 2015). Researchers could provide insights on how uncertainty could be mitigated and transformed into opportunities, for example, through the lens of digital innovations (e.g., Nylén and Holmström, 2015), entrepreneurship theories (e.g., Read and Sarasvathy 2012; Nambisan 2017) or design thinking (e.g., Luotola et al., 2017).

Digital data provides novel forms of behavioral targeting for financial or other motives; however, the use of *data as a capital* impacts the liberty, autonomy, and well-being of individuals (Zuboff, 2015). There is an increasing concern related to surveillance abuses by governments where states might "govern by identity" (Whitley et al., 2014). SSI ecosystems provide a new aspect to the discussions related to the data as a capital and as a means for surveillance. Thus, researchers might further investigate these complex policy, legal and ethical issues considering the potential of SSI on fighting social injustice caused by asymmetrical personal data accumulation and analytics (Cinnamon, 2017). Privacy concerns, and the War of Wallets, could be studied from the viewpoint of the logic of accumulation in the surveillance economy (e.g., Zuboff, 2015; Cohen, 2018).

## Conclusions

Based on this study, we define SSI ecosystems as innovation ecosystems describing the collaborative effort of a diverse set of actors towards innovation, such as (1) *the infrastructure providers* delivering a technology and governance dual framework and (2) *various partner companies and end customers* that provide complementary products and services as well as demands and capabilities. In SSI ecosystems, the trust among actors is established through peer-to-peer interactions between the *issuers*, *verifiers* and *holders* (i.e., the trust triangle). SSI solutions require *conceptualization*, such as defining the purpose, mission, context and boundary conditions. They consist of *technology and governance architectures* including the technological components and the business, legal and technical rules and policies, and these two frameworks should be developed hand in hand. SSI systems require orchestrating a *collaborative ecosystem* consisting of a set of actors with diverse, sometimes conflicting incentives (see, e.g., De Filippi and Loveluck, 2016). Their adoption requires fundamental changes in existing processes, new *business models*, and forming of new strategic alliances (see, e.g., Chen and Bellavitis, 2020).

Several factors affect the characteristics, adoption and orchestration of SSI ecosystems. They reside in a highly uncertain environment characterized by high volatility, fast changes, and information asymmetry. The SSI market is fragmented, the standards are yet immature, and there is a legal and regulatory uncertainty. SSI ecosystems are strongly affected by the cultural and country characteristics that they reside in. Furthermore, the industrial characteristics shape their requirements as well as their perceived benefits and sacrifices. Finally, the organizational culture, the actors' individual characteristics, and the expert communities, non-profit organizations and standard setting organizations play an important role in their structure and adoption.

In certain cases, SSI holds the promise to provide trustful, simple, safe and private digital interactions. They facilitate innovations, cost savings and increase in job performance (Zwitter et al., 2020). They provide means to counteract money laundering, fraud and other economic crimes (Evernym, 2020). They might also provide an opportunity to hinder the oligopoly structure of today's Internet where digital identities and personal data are primarily managed by the "Tech Giants" (Der et al., 2017).They facilitate the formation of new strategic alliances and new forms of joint actions. However, several challenges are associated with their development and adoption, such as, technical immaturity, the lack of interoperability and standards, challenges related to ecosystem governance, lack of knowledge, information asymmetry, funding and legal issues, and the tension between the allocation of citizenship and "innate" individual rights (e.g., Prewett et al., 2019, Gstrein and Kochenov, 2020).

Being a nascent technology, with the promise of building a trust layer into digital interactions, SSI and the ecosystems built around it claim for further studies. In particular, there is a need for further research related to the building blocks, adoption, orchestration and operation of SSI ecosystems. Some studies assess different SSI solutions (e.g., Naik and Jenkins, 2020); however, we lack work both from practitioners and researchers on the technical and governance requirements, the architecture, design principles and the interoperability, among others. In this paper, we propose further research avenues on different domains, such as digital trust, digital identity, identity management, ecosystems, digital innovations, value co-creation, open source communities and standard setting organizations, digital inclusion, and value-based selling.

The limitations of the study are rooted in the limitations of the qualitative research approach. Further studies might investigate this domain by using other research methods, such as statistical methods, analytical methods or system dynamics. Furthermore, we acknowledge that the topic is rather new, developing fast and this might impact the validity of the results. To mitigate the possibility of this bias, we based our results on up-to-date sources both from scientific and grey literature in order to provide a state-of-the-art and state-of-practice overview of the SSI ecosystems.

**Acknowledgment.** This research has been conducted in the APPIA research project in the Security And Software Engineering Research Center (S^2 ERC, 2020-21) and funded by Business Finland. We are grateful for all experts who contributed to this research by sharing their insights.